\documentclass[aps,preprint]{revtex4}%
\usepackage{amsfonts}
\usepackage{amsmath}
\usepackage{amssymb}
\usepackage{graphicx}%
\providecommand{\U}[1]{\protect\rule{.1in}{.1in}}
\begin{document}
\title[ ]{Classical Electromagnetic Interaction of a Charge with a Solenoid or Toroid}
\author{Timothy H. Boyer}
\affiliation{Department of Physics, City College of the City University of New York, New
York, New York 10031, USA}

\begin{abstract}
The Aharonov-Bohm phase shift in a particle interference pattern when
electrons pass a long solenoid is identical in form with the optical
interference pattern shift when a piece of retarding glass is introduced into
one path of a two-beam optical interference pattern. \ The particle
interference-pattern deflection is a relativistic effect of order $1/c^{2}$,
though this \textit{relativity} aspect is rarely mentioned in the literature.
\ Here we give a thorough analysis of the classical electromagnetic aspects of
the interaction between a solenoid or toroid and a charged particle. \ We
point out the magnetic Lorentz force which the solenoid or toroid experiences
due to a passing charge. \ Although analysis in the rest frame of the solenoid
or toroid will involve back Faraday fields on the charge, the analysis in the
inertial frame in which the charge is initially at rest involves forces due to
only \textit{electric} fields where forces are equal in magnitude and opposite
in direction. \ The classical analysis is made using the Darwin Lagrangian.
\ We point out that the classical analysis suggests an angular deflection
independent of Planck's constant $\hbar$,\ where the deflection magnitude is
identical with that given by the traditional quantum analysis, but where the
deflection direction is unambiguous. \ 

\end{abstract}
\maketitle

\section{Introduction}

\subsection{Optical Phase Shift Analogous to the Aharonov-Bohm Deflection}

If the optical beam from a small source is split into two beams and then is
recombined, it will, in general, produce an interference pattern (dependent
upon the difference in optical path length between the two beams) inside a
single-beam envelope (dependent upon path-length differences within each
individual beam). \ This optical situation is familiar for two slits in a
barrier producing an optical interference pattern on a distant screen. \ If a
piece of glass is introduced into just one to the two beams, then there will
be a deflection of the double-slit interference pattern involving both beams
but not in the single-slit envelope which involves each beam separately. \ The
direction of deflection of the optical beam is unambiguously toward the side
on which the piece of glass slowed the speed of the beam. \ This deflection in
the optical double-beam interference pattern takes exactly the same form as
the magnetic Aharonov-Bohm deflection of the double-beam interference pattern
involving two beams from a charged-particle source which are separated, pass
around a long solenoid, and then are recombined. \ 

\subsection{The Magnetic Aharonov-Bohm Situation}

When electrons pass around the sides of a long solenoid, they produce an
interference pattern which changes as the solenoid magnetic field is
altered.\cite{M1962}\cite{Bayh1962}\ \ A figure in the experimental
report\cite{Bayh1962} of this effect shows clearly the deflection of the
double-beam interference pattern with the undisplaced single-beam envelope as
the current in the solenoid is increased. \ This Aharonov-Bohm\cite{AB1959}
phase shift is often claimed to have no basis in classical electrodynamics
because the electrons move in a region where the magnetic fields of an
\textit{unperturbed} infinite solenoid vanish. \ It is the magnetic flux which
is said to affect the charges' behavior. \ Many physicists have accepted the
claim that the Aharonov-Bohm phase shift represents an effect of the vector
potential which has no analogue in classical electrodynamics.\cite{F1964}%
\cite{Shad}\cite{Garg}\cite{Gq}\cite{Ball} \ However, when asked in which
direction the double-slit interference pattern is deflected, many physicists
draw a blank. \ Indeed, although the theoretical and experimental literature
on the Aharonov-Bohm is voluminous and the \textit{magnitude} of the phase
shift is confirmed experimentally, it seems very hard to find an authoritative
suggestion of the direction of the deflection in either quantum theory or in
the reports of the experiments. \ Furthermore, the naive discussion of
magnetic energy changes suggests a relative particle lag (and hence
deflection) which is in the opposite direction from that suggested by a
detailed treatment of the classical electrodynamic situation. \ If the
magnetic field of a solenoid is upwards, and charges pass around the sides of
the solenoid, is the Aharonov-Bohm deflection in the same direction or in the
opposite direction compared with the Lorentz-force deflection of the charge
when passing through the center of the solenoid? \ And has this direction been
confirmed experimentally? \ 

\subsection{Classical Electromagnetic Analysis}

The Aharonov-Bohm deflection involves charged particles and solenoids which
are traditional aspects of classical electromagnetic theory. \ Thus it seems
natural to discuss the interaction of a solenoid or toroid with a passing
charge from the viewpoint of classical electrodynamics. \ The present article
is the research-oriented discussion in a series of four articles. \ The first
article pointed out that unsymmetric magnetic energy changes are unfamiliar to
many physicists who have seen only the highly symmetric magnetic energy
changes of the textbook examples.\cite{Forces} \ The second article treated
the Aharonov-Bohm interaction as a relativity paradox in the style of the
pole-in-the-barn paradox where analysis is easier in one inertial frame than
another.\cite{Paradox} \ A short third article considered simply the direction
of the double-slit deflection.\cite{direction} \ The full, careful analysis is
the subject of the present article. \ 

\subsection{Outline of the Present Article}

In the present article, we give a step-by-step analysis going from familiar
electrostatics to unfamiliar aspects of classical electrodynamics. \ We start
with the electrostatic interaction of a point charge passing down the axis of
symmetry of a circular electric line charge, and end with a point charge
passing down the symmetry axis of a magnetic toroid. \ First, we introduce the
Darwin Lagrangian\cite{Jackson} which is appropriate for quasistatic
electromagnetic systems, where radiation is excluded. \ Such a
relativistic-to-order-$1/c^{2}$ interaction requires the use of point charges.
\ Secondly, it is pointed out that a circular ring of many equally-spaced
point charges in close proximity exhibits a large \textit{electromagnetic}
inertia which tends to prevent changes in the motions of the discrete charges
in the ring. \ Thirdly, this point of view is applied to an
\textit{electrostatic analogue} of the magnetic Aharonov-Bohm situation where,
in the analogue, the charge passes through a ring of electric dipoles.
Fourthly, the magnetic Aharonov-Bohm situation is discussed, and the forces
are identified. \ The magnitude of the resulting classical electromagnetic lag
for the passing charge is exactly what was suggested fifty years
ago,\cite{B1973c} and is precisely what is needed to account for the observed
interference pattern shift. \ The classical electromagnetic lag depends on the
magnetic field in the toroid and so on the integral of the vector potential
around the toroid. \ The phase shift when a solenoid is placed between the two
holes of a double-hole mask involves the ratio of the classical lag to the
double-hole separation. \ The angular deflection associated with the lag (or
the fractional phase shift) is independent of Planck's constant $\hbar$. \ The
direction of deflection is identified unambiguously in classical theory.
\ Fifthly, we place the Aharonov-Bohm phase shift in the context of poorly
understood interactions between charges and magnets. \ We point out some
historical aspects regarding the understanding of the phase shift, and we note
the possibility of further experimental exploration.

\section{The Darwin Lagrangian}

The experimental observations of the interaction of a point charge and a
solenoid generally involve relatively slowly-moving charges, both in the wires
of the solenoid and for the passing charges. \ The natural classical
electromagnetic approximation for the situation is given by the Darwin
Lagrangian,\cite{Jackson}%
\begin{align}
L  &  =%
%TCIMACRO{\tsum \nolimits_{a}}%
%BeginExpansion
{\textstyle\sum\nolimits_{a}}
%EndExpansion
\left\{  -M_{a}c^{2}\sqrt{1-\left(  \mathbf{\dot{r}}_{a}\mathbf{/}c\right)
^{2}}\right\}  -\frac{1}{2}%
%TCIMACRO{\tsum \nolimits_{a}}%
%BeginExpansion
{\textstyle\sum\nolimits_{a}}
%EndExpansion%
%TCIMACRO{\tsum \nolimits_{b\neq a}}%
%BeginExpansion
{\textstyle\sum\nolimits_{b\neq a}}
%EndExpansion
\frac{q_{a}q_{b}}{\left\vert \mathbf{r}_{a}-\mathbf{r}_{b}\right\vert
}\nonumber\\
&  +\frac{1}{2}%
%TCIMACRO{\tsum \nolimits_{a}}%
%BeginExpansion
{\textstyle\sum\nolimits_{a}}
%EndExpansion%
%TCIMACRO{\tsum \nolimits_{b\neq a}}%
%BeginExpansion
{\textstyle\sum\nolimits_{b\neq a}}
%EndExpansion
\left\{  \frac{q_{a}q_{b}}{2c^{2}}\left(  \frac{\mathbf{\dot{r}}%
_{a}\mathbf{\cdot\dot{r}}_{b}}{\left\vert \mathbf{r}_{a}-\mathbf{r}%
_{b}\right\vert }+\frac{\left[  \mathbf{\dot{r}}_{a}\mathbf{\cdot}\left(
\mathbf{r}_{a}-\mathbf{r}_{b}\right)  \right]  \left[  \mathbf{\dot{r}}%
_{b}\cdot\left(  \mathbf{r}_{a}-\mathbf{r}_{b}\right)  \right]  }{\left\vert
\mathbf{r}_{a}-\mathbf{r}_{b}\right\vert ^{3}}\right)  \right\}  , \label{DL}%
\end{align}
where the square root $\sqrt{1-\left(  \mathbf{\dot{r}}_{a}\mathbf{/}c\right)
^{2}}$ should be expanded through order $\left(  \mathbf{\dot{r}}%
_{a}\mathbf{/}c\right)  ^{4}$. \ The Euler-Lagrange equations of motion are
\begin{equation}
\frac{d}{dt}\left(  \frac{\partial L}{\partial\mathbf{\dot{r}}_{a}}\right)
-\frac{\partial L}{\partial\mathbf{r}_{a}}=0,
\end{equation}
where it is understood that each rectangular component of the vectors
$\mathbf{r}_{a}$ and $\mathbf{\dot{r}}_{a}$ is to be treated separately. \ The
total energy of the system is given by
\begin{align}
U  &  =%
%TCIMACRO{\tsum \nolimits_{a}}%
%BeginExpansion
{\textstyle\sum\nolimits_{a}}
%EndExpansion
\frac{M_{a}c^{2}}{\sqrt{1-\left(  \mathbf{\dot{r}}_{a}\mathbf{/}c\right)
^{2}}}+\frac{1}{2}%
%TCIMACRO{\tsum \nolimits_{a}}%
%BeginExpansion
{\textstyle\sum\nolimits_{a}}
%EndExpansion%
%TCIMACRO{\tsum \nolimits_{b\neq a}}%
%BeginExpansion
{\textstyle\sum\nolimits_{b\neq a}}
%EndExpansion
q_{a}\Phi_{b}\left(  \mathbf{r}_{a},t\right)  +\frac{1}{2}%
%TCIMACRO{\tsum \nolimits_{a}}%
%BeginExpansion
{\textstyle\sum\nolimits_{a}}
%EndExpansion%
%TCIMACRO{\tsum \nolimits_{b\neq a}}%
%BeginExpansion
{\textstyle\sum\nolimits_{b\neq a}}
%EndExpansion
q_{a}\frac{\mathbf{v}_{a}}{c}\cdot\mathbf{A}_{b}\left(  \mathbf{r}%
_{a},t\right) \nonumber\\
&  =%
%TCIMACRO{\tsum \nolimits_{a}}%
%BeginExpansion
{\textstyle\sum\nolimits_{a}}
%EndExpansion
\frac{M_{a}c^{2}}{\sqrt{1-\left(  \mathbf{\dot{r}}_{a}\mathbf{/}c\right)
^{2}}}+\frac{1}{2}%
%TCIMACRO{\tsum \nolimits_{a}}%
%BeginExpansion
{\textstyle\sum\nolimits_{a}}
%EndExpansion%
%TCIMACRO{\tsum \nolimits_{b\neq a}}%
%BeginExpansion
{\textstyle\sum\nolimits_{b\neq a}}
%EndExpansion
\frac{q_{a}q_{b}}{\left\vert \mathbf{r}_{a}-\mathbf{r}_{b}\right\vert
}\nonumber\\
&  +\left(  1/2\right)
%TCIMACRO{\tsum \nolimits_{a}}%
%BeginExpansion
{\textstyle\sum\nolimits_{a}}
%EndExpansion%
%TCIMACRO{\tsum \nolimits_{b\neq a}}%
%BeginExpansion
{\textstyle\sum\nolimits_{b\neq a}}
%EndExpansion
\left\{  \frac{q_{a}q_{b}}{2c^{2}}\right\}  \left\{  \frac{\mathbf{\dot{r}%
}_{a}\mathbf{\cdot\dot{r}}_{b}}{\left\vert \mathbf{r}_{a}-\mathbf{r}%
_{b}\right\vert }+\frac{\left[  \mathbf{\dot{r}}_{a}\mathbf{\cdot}\left(
\mathbf{r}_{a}-\mathbf{r}_{b}\right)  \right]  \left[  \mathbf{\dot{r}}%
_{b}\cdot\left(  \mathbf{r}_{a}-\mathbf{r}_{b}\right)  \right]  }{\left\vert
\mathbf{r}_{a}-\mathbf{r}_{b}\right\vert ^{3}}\right\}  ,
\end{align}
including the particle mechanical energy involving the masses $M_{a}$, the
energy in the electric field involving separations between the charges, and
the energy in the magnetic field involving the velocities $\mathbf{\dot{r}%
}_{a}$ of the charges. \ The associated total linear momentum for the system
is
\begin{align}
\mathbf{P}  &  \mathbf{=}%
%TCIMACRO{\tsum \nolimits_{a}}%
%BeginExpansion
{\textstyle\sum\nolimits_{a}}
%EndExpansion
\left(  \frac{M_{a}\mathbf{\dot{r}}_{a}}{\sqrt{1-\left(  \mathbf{\dot{r}}%
_{a}\mathbf{/}c\right)  ^{2}}}+%
%TCIMACRO{\tsum \nolimits_{b\neq a}}%
%BeginExpansion
{\textstyle\sum\nolimits_{b\neq a}}
%EndExpansion
\frac{q_{a}}{c}\mathbf{A}_{b}\left(  \mathbf{r}_{a},t\right)  \right)
\nonumber\\
&  \mathbf{=}%
%TCIMACRO{\tsum \nolimits_{a}}%
%BeginExpansion
{\textstyle\sum\nolimits_{a}}
%EndExpansion
\frac{M_{a}\mathbf{\dot{r}}_{a}}{\sqrt{1-\left(  \mathbf{\dot{r}}%
_{a}\mathbf{/}c\right)  ^{2}}}+%
%TCIMACRO{\tsum \nolimits_{a}}%
%BeginExpansion
{\textstyle\sum\nolimits_{a}}
%EndExpansion%
%TCIMACRO{\tsum \nolimits_{b\neq a}}%
%BeginExpansion
{\textstyle\sum\nolimits_{b\neq a}}
%EndExpansion
q_{a}\frac{q_{b}}{2c^{2}}\left(  \frac{\mathbf{\dot{r}}_{b}}{\left\vert
\mathbf{r}_{a}-\mathbf{r}_{b}\right\vert }+\frac{\left(  \mathbf{r}%
_{a}-\mathbf{r}_{b}\right)  \left[  \mathbf{\dot{r}}_{b}\cdot\left(
\mathbf{r}_{a}-\mathbf{r}_{b}\right)  \right]  }{\left\vert \mathbf{r}%
_{a}-\mathbf{r}_{b}\right\vert ^{3}}\right)  ,
\end{align}
and includes both mechanical linear momentum and electromagnetic field
momentum. \ The electromagnetic potentials are given by a scalar potential
\begin{equation}
V(\mathbf{r},t)=%
%TCIMACRO{\tsum \nolimits_{a}}%
%BeginExpansion
{\textstyle\sum\nolimits_{a}}
%EndExpansion
V_{a}\left(  \mathbf{r,}t\right)  =%
%TCIMACRO{\tsum \nolimits_{a}}%
%BeginExpansion
{\textstyle\sum\nolimits_{a}}
%EndExpansion
\frac{q_{a}}{\left\vert \mathbf{r}-\mathbf{r}_{a}\right\vert },
\end{equation}
and a vector potential
\begin{equation}
\mathbf{A}\left(  \mathbf{r,}t\right)  =%
%TCIMACRO{\tsum \nolimits_{a}}%
%BeginExpansion
{\textstyle\sum\nolimits_{a}}
%EndExpansion
\mathbf{A}_{a}\left(  \mathbf{r,}t\right)  =%
%TCIMACRO{\tsum \nolimits_{a}}%
%BeginExpansion
{\textstyle\sum\nolimits_{a}}
%EndExpansion
\frac{q_{a}}{2c}\left(  \frac{\mathbf{\dot{r}}_{a}}{\left\vert \mathbf{r}%
-\mathbf{r}_{a}\right\vert }+\frac{\left(  \mathbf{r}-\mathbf{r}_{a}\right)
\left[  \mathbf{\dot{r}}_{a}\cdot\left(  \mathbf{r}-\mathbf{r}_{a}\right)
\right]  }{\left\vert \mathbf{r}-\mathbf{r}_{a}\right\vert ^{3}}\right)  .
\end{equation}
If the Euler-Lagrange equations of motion are rewritten in terms of electric
and magnetic fields in the Lorentz force, then one obtains%
\begin{align}
\frac{d}{dt}\left(  \frac{M_{a}\mathbf{\dot{r}}_{a}}{\sqrt{1-\left(
\mathbf{\dot{r}}_{a}\mathbf{/}c\right)  ^{2}}}\right)   &  =q_{a}%
\mathbf{E}\left(  \mathbf{r}_{a},t\right)  +q_{a}\frac{\mathbf{\dot{r}}_{a}%
}{c}\times\mathbf{B}\left(  \mathbf{r}_{a},t\right) \nonumber\\
&  =q_{a}%
%TCIMACRO{\tsum \nolimits_{b\neq a}}%
%BeginExpansion
{\textstyle\sum\nolimits_{b\neq a}}
%EndExpansion
\left(  -\nabla_{a}V_{b}(\mathbf{r}_{a},t)-\frac{1}{c}\frac{\partial
\mathbf{A}_{b}(\mathbf{r}_{a},t)}{\partial t}\right) \nonumber\\
&  +q_{a}\frac{\mathbf{\dot{r}}_{a}}{c}\times\left[  \nabla_{a}\times%
%TCIMACRO{\tsum \nolimits_{b\neq a}}%
%BeginExpansion
{\textstyle\sum\nolimits_{b\neq a}}
%EndExpansion
\mathbf{A}_{b}(\mathbf{r}_{a},t)\right]  ,
\end{align}
where
\begin{align}
\mathbf{E}\left(  \mathbf{r,}t\right)   &  =%
%TCIMACRO{\tsum \nolimits_{a}}%
%BeginExpansion
{\textstyle\sum\nolimits_{a}}
%EndExpansion
\frac{q_{a}\left(  \mathbf{r}-\mathbf{r}_{a}\right)  }{\left\vert
\mathbf{r}-\mathbf{r}_{a}\right\vert ^{3}}\left[  1+\frac{1}{2}\left(
\frac{\mathbf{\dot{r}}_{a}}{c}\right)  ^{2}-\frac{3}{2}\left(  \frac
{\mathbf{\dot{r}}_{a}\cdot\left(  \mathbf{r}-\mathbf{r}_{a}\right)
}{c\left\vert \mathbf{r}-\mathbf{r}_{a}\right\vert }\right)  ^{2}\right]
\nonumber\\
&  -%
%TCIMACRO{\tsum \nolimits_{a}}%
%BeginExpansion
{\textstyle\sum\nolimits_{a}}
%EndExpansion
\frac{q_{a}}{2c^{2}}\left[  \frac{\mathbf{\ddot{r}}_{a}}{\left\vert
\mathbf{r}-\mathbf{r}_{a}\right\vert }+\frac{\left(  \mathbf{r}-\mathbf{r}%
_{a}\right)  \left[  \mathbf{\ddot{r}}_{a}\cdot\left(  \mathbf{r}%
-\mathbf{r}_{a}\right)  \right]  }{\left\vert \mathbf{r}-\mathbf{r}%
_{a}\right\vert ^{3}}\right]  \label{Ert}%
\end{align}
and
\begin{equation}
\mathbf{B}\left(  \mathbf{r}\,,t\right)  =%
%TCIMACRO{\tsum \nolimits_{a}}%
%BeginExpansion
{\textstyle\sum\nolimits_{a}}
%EndExpansion
q_{a}\frac{\mathbf{\dot{r}}_{a}}{c}\times\frac{\left(  \mathbf{r}%
-\mathbf{r}_{a}\right)  }{\left\vert \mathbf{r}-\mathbf{r}_{a}\right\vert
^{3}}.
\end{equation}
We draw attention particularly to the acceleration terms\cite{PA} involving
$\mathbf{\ddot{r}}_{a}$ in the electric field in Eq. (\ref{Ert}). \ It is
these terms which are associated with back (Faraday) electric fields whose
work balances changes in magnetic energies involving changes of speed for
charged particles. \ The Darwin Lagrangian in Eq. (\ref{DL}) conserves energy,
linear momentum, and angular momentum. \ Constant motion of the center of
energy holds only through order $1/c^{2}$.

\section{Electromagnetic Inertia of a Ring of Charges}

\subsection{Axial Symmetry}

Although the Aharonov-Bohm interaction is usually discussed with solenoids, it
is far easier to deal with toroids. \ Thus we will be interested in the
classical electromagnetic interaction of a charge $e$ with various systems,
which systems (for ease of calculation) all involve axial symmetry. \ We take
the axis of symmetry as the $z$-axis. \ In our discussions before dealing with
a magnetic toroid, all motions are parallel to the $z$-axis, and there are
forces of constraint which prevent any motions perpendicular to the $z$-axis. \ 

\subsection{Nonrelativistic Interaction of a Charge $e$ and a Circular Line
Charge $\lambda$}

We start with the \textit{nonrelativistic} interaction of a particle of charge
$e$ and mass $M_{e}$, located on the $z$-axis at $z_{e}$, with a uniform
circular line charge of radius $R$ and total charge $q$ (with charge per unit
length $\lambda=q/\left(  2\pi R\right)  $), and mass $M_{q}$, with center
located on the $z$-axis at $z_{q}$. \ The plane of the circular line charge is
parallel to the $xy$-plane. \ Then, in the nonrelativistic analysis, the
charges $e$ and $q$ experience equal-and-opposite forces and have
accelerations $a_{e}$ and $a_{q}$ parallel to the $z$-axis given by%
\begin{equation}
a_{e}=\frac{eq\left(  z_{e}-z_{q}\right)  }{M_{e}\left\vert \left(
z_{e}-z_{q}\right)  ^{2}+R^{2}\right\vert ^{3/2}}\text{ \ \ and \ \ }%
a_{q}=\frac{qe\left(  z_{q}-z_{e}\right)  }{M_{q}\left\vert \left(
z_{q}-z_{e}\right)  ^{2}+R^{2}\right\vert ^{3/2}}. \label{aNR}%
\end{equation}

\subsection{Relativistic Interaction Through Order $1/c^{2}$}

The Darwin Lagrangian (which gives relativistic electromagnetic effects
through order $1/c^{2}$) requires that we deal not with continuous charge
distributions but rather with discrete point charges. \ Thus we imagine
replacing the circular line charge $\lambda$ by $N$ individual point charges
$q_{i}=q/N$ of mass $M_{i}=M_{q}/N,$ spaced uniformly around the circle of
radius $R$. \ We imagine that each of the charges is allowed to move along its
own frictionless rod parallel to the $z$-axis. \ The rods provide forces of
constraint perpendicular to the $z$-axis but introduce no energy.

The relativistic analysis using the Darwin Lagrangian leads to equations of
motion parallel to the $z$-axis which include the interactions between the
point charges $q_{i}$ and $q_{j}$, both of which are in the ring of charges.
\ Thus, for small particle speeds $v<<c$, we find from the electric field in
Eq. (\ref{Ert})
\begin{align}
M_{e}a_{e}  &  =e%
%TCIMACRO{\tsum \nolimits_{i}}%
%BeginExpansion
{\textstyle\sum\nolimits_{i}}
%EndExpansion
q_{i}\frac{\left(  z_{e}-z_{i}\right)  }{\left\vert \mathbf{r}_{e}%
-\mathbf{r}_{i}\right\vert ^{3}}-e%
%TCIMACRO{\tsum \nolimits_{i}}%
%BeginExpansion
{\textstyle\sum\nolimits_{i}}
%EndExpansion
\frac{q_{i}}{2c^{2}}\left[  \frac{a_{i}}{\left\vert \mathbf{r}_{e}%
-\mathbf{r}_{i}\right\vert }+\frac{a_{i}\left[  \left(  z_{e}-z_{i}\right)
^{2}\right]  }{\left\vert \mathbf{r}_{e}-\mathbf{r}_{i}\right\vert ^{3}%
}\right] \nonumber\\
&  =eq\frac{\left(  z_{e}-z_{q}\right)  }{\left\vert \left(  z_{e}%
-z_{q}\right)  ^{2}+R^{2}\right\vert ^{3/2}}-\frac{eq}{2c^{2}}\left[
\frac{a_{q}}{\left\vert \left(  z_{e}-z_{q}\right)  ^{2}+R^{2}\right\vert
^{1/2}}+\frac{a_{q}\left(  z_{e}-z_{q}\right)  ^{2}}{\left\vert \left(
z_{e}-z_{q}\right)  ^{2}+R^{2}\right\vert ^{3/2}}\right]  \label{aee}%
\end{align}
and%
\begin{align}
M_{i}a_{q}  &  =q_{i}\frac{e\left(  z_{i}-z_{e}\right)  }{\left\vert
\mathbf{r}_{i}-\mathbf{r}_{e}\right\vert ^{3}}-q_{i}\frac{e}{2c^{2}}\left\{
\frac{a_{e}}{\left\vert \mathbf{r}_{i}-\mathbf{r}_{e}\right\vert }+\frac
{a_{e}\left(  z_{i}-z_{e}\right)  ^{2}}{\left\vert \mathbf{r}_{i}%
-\mathbf{r}_{e}\right\vert ^{3}}\right\} \nonumber\\
&  -q_{i}%
%TCIMACRO{\tsum \nolimits_{j\neq i}}%
%BeginExpansion
{\textstyle\sum\nolimits_{j\neq i}}
%EndExpansion
\frac{q_{j}}{2c^{2}}\frac{a_{j}}{\left\vert \mathbf{r}_{i}-\mathbf{r}%
_{j}\right\vert }\nonumber\\
&  =\frac{q}{N}\frac{e\left(  z_{q}-z_{e}\right)  }{\left\vert \left(
z_{q}-z_{e}\right)  ^{2}+R^{2}\right\vert ^{3/2}}-\frac{q}{N}\frac{e}{2c^{2}%
}\left\{  \frac{a_{e}}{\left\vert \left(  z_{q}-z_{e}\right)  ^{2}%
+R^{2}\right\vert ^{1/2}}+\frac{a_{e}\left(  z_{q}-z_{e}\right)  ^{2}%
}{\left\vert \left(  z_{q}-z_{e}\right)  ^{2}+R^{2}\right\vert ^{3/2}}\right\}
\nonumber\\
&  -\frac{q}{N}%
%TCIMACRO{\tsum \nolimits_{j\neq i}}%
%BeginExpansion
{\textstyle\sum\nolimits_{j\neq i}}
%EndExpansion
\frac{q/N}{2c^{2}}\frac{a_{q}}{\left\vert \mathbf{r}_{i}-\mathbf{r}%
_{j}\right\vert } \label{aq}%
\end{align}
Here we have used the fact that (because of the axial symmetry) all of the
charges $q_{i}$ will have the same acceleration $a_{i}=a_{q}$. \ However, the
situation in Eqs. (\ref{aee}) and (\ref{aq}) is completely different from the
nonrelativistic equations in Eq. \ref{aNR}). \ 

In order to account for changes in magnetic field energy, the accelerations
$a_{e}$ and $a_{q}$ of the charge $e$ and of the ring of charges $q_{i}$ now
appear on \textit{both} sides of the equations of motion. \ \ The equations
for the accelerations $a_{e}$ and $a_{q}$ are coupled. \ The accelerations for
the charge $e$ and the ring $q$ can be determined only after solving the
coupled equations. \ The equations can be rewritten as
\begin{equation}
\frac{eq\left(  z_{e}-z_{q}\right)  }{\left\vert \left(  z_{e}-z_{q}\right)
^{2}+R^{2}\right\vert ^{3/2}}=M_{e}a_{e}+\frac{eq}{2c^{2}}\left[  \frac
{1}{\left\vert \left(  z_{e}-z_{q}\right)  ^{2}+R^{2}\right\vert ^{1/2}}%
+\frac{\left(  z_{e}-z_{q}\right)  ^{2}}{\left\vert \left(  z_{e}%
-z_{q}\right)  ^{2}+R^{2}\right\vert ^{3/2}}\right]  a_{q} \label{FDar1}%
\end{equation}
and%
\begin{align}
\frac{q}{N}\frac{e\left(  z_{q}-z_{e}\right)  }{\left\vert \left(  z_{q}%
-z_{e}\right)  ^{2}+R^{2}\right\vert ^{3/2}}  &  =\left[  \frac{eq}{N2c^{2}%
}\left\{  \frac{1}{\left\vert \left(  z_{q}-z_{e}\right)  ^{2}+R^{2}%
\right\vert ^{1/2}}+\frac{\left(  z_{q}-z_{e}\right)  ^{2}}{\left\vert \left(
z_{q}-z_{e}\right)  ^{2}+R^{2}\right\vert ^{3/2}}\right\}  \right]
a_{e}\nonumber\\
&  +\left[  \frac{M_{q}}{N}+\left(  \frac{q}{N}\right)  ^{2}\frac{1}{2c^{2}}%
%TCIMACRO{\tsum \nolimits_{j\neq1}}%
%BeginExpansion
{\textstyle\sum\nolimits_{j\neq1}}
%EndExpansion
\frac{1}{\left\vert \mathbf{r}_{1}-\mathbf{r}_{j}\right\vert }\right]  a_{q}.
\label{FDar2}%
\end{align}
Except for a factor of $N$, the electrostatic forces on the left-hand side of
Eqs. (\ref{FDar1}) and (\ref{FDar2}) are equal in magnitude and opposite in
direction. \ The right-hand sides correspond to changes in momentum, including
both particle momentum and electromagnetic field momentum. \ 

\subsection{Electromagnetic Inertia of Closely-Spaced Point Charges}

In the last line of Eq. (\ref{FDar2}), it is crucial to note that the sum over
the charges $j\neq1$ acts like an electromagnetic inertial contribution in
addition to the mass $M_{q}/N$ of the charge $q_{i}$ in the ring. \ As the
number $N$ of charges in the ring increases while the radius $R$ of the ring
remains fixed, the electromagnetic inertial contribution increases and
eventually dominates the mass $M_{q}$, since $%
%TCIMACRO{\tsum \nolimits_{j=2}^{N}}%
%BeginExpansion
{\textstyle\sum\nolimits_{j=2}^{N}}
%EndExpansion
\left(  1/\left\vert \mathbf{r}_{1}-\mathbf{r}_{j}\right\vert \right)  $
increases as $N^{2}$ for large $N~$\ and fixed radius $R$. \ As the number
$N$\ of charges $q_{i}$ in the ring increases, the separations $\left\vert
\mathbf{r}_{i}-\mathbf{r}_{j}\right\vert $ between the charges $q_{i}$ and
$q_{j}$ becomes ever smaller and the number of charges per unit length
contributing to the back field also increases. \ Therefore the influence of
the back (Faraday) acceleration terms becomes ever larger, while the actual
acceleration of each charge $q_{i}$ becomes ever smaller. \ The charge $e$ is
some distance away from the ring, and so $e$\ does not experience an
increasing contribution with $N$ from the charges $q_{i}$ all of which are at
the same distance from $e$. \ 

\subsection{Simplification in the Large-Electromagnetic-Inertia Limit}

For large numbers $N$ of closely-spaced particles $q_{i}$ in the ring of
charges, the electromagnetic inertia of the ring can become very large and the
acceleration $a_{q}$ very small. \ Therefore in Eq. (\ref{FDar1}) the term in
$a_{q}$ can be neglected. \ But then the acceleration $a_{e}/c^{2}$ is small
and can be neglected in Eq. (\ref{FDar2}). \ The mechanical mass $M_{q}$ of
the ring of particles is negligible compared to the electromagnetic terms when
$N$ is large. \ Therefore, when the second equation is multiplied by $N$, the
equations (\ref{FDar1}) and (\ref{FDar2}) simplify to
\begin{equation}
\frac{eq\left(  z_{e}-z_{q}\right)  }{\left\vert \left(  z_{e}-z_{q}\right)
^{2}+R^{2}\right\vert ^{3/2}}=M_{e}a_{e} \label{eqmea}%
\end{equation}
and
\begin{equation}
\frac{qe\left(  z_{q}-z_{e}\right)  }{\left\vert \left(  z_{q}-z_{e}\right)
^{2}+R^{2}\right\vert ^{3/2}}=\left(  \frac{q^{2}}{2c^{2}}\frac{1}{N}%
%TCIMACRO{\tsum \nolimits_{j\neq1}}%
%BeginExpansion
{\textstyle\sum\nolimits_{j\neq1}}
%EndExpansion
\frac{1}{\left\vert \mathbf{r}_{1}-\mathbf{r}_{j}\right\vert }\right)  a_{q}.
\label{qeMa}%
\end{equation}
\ 

In this situation of large electromagnetic inertia for the ring, the external
charge $e$ behaves as though it were interacting with a very massive ring of
total charge $q$. \ The electromagnetic inertia arising from the (Faraday)
acceleration terms within the ring acts to prevent changes in the velocities
of the closely-spaced charges. \ Without the presence of the other charges,
each charge $q_{i}$ of small mass $M_{q}/N$ would have a large acceleration
due to the external charge $e$. \ However, this potential large acceleration
of each particle $q_{i}$ is counteracted by the acceleration electric field of
the other charges of the ring. \ The \textit{mechanical} linear momentum
change of the charge $e$ in Eq. (\ref{eqmea}) is balanced against the
\textit{electromagnetic} linear momentum change of the ring of charges in Eq.
(\ref{qeMa}). \ 

\subsection{Failure of the $1/c$ Hierarchy}

We notice that when the electromagnetic inertia is large, the comparison of
the number of factors of $1/c$ is no longer helpful. \ Thus each term
contributing to the \textit{electromagnetic} inertia is of order $1/c^{2}$
whereas the \textit{particle} inertial masses $M_{e}$ and $M_{q}$ are zero
order in $1/c$. \ This transition in orders of $1/c$ (where the
electromagnetic inertia dominates the mechanical inertia) is most familiar
when dealing with the magnetic energy of a solenoid. \ The magnetic energy is
of order $1/c^{2}$ whereas the kinetic energy of the charge carriers in a
solenoid is zero order in $1/c$. \ However, the electromagnetic energy in a
solenoid is so much larger than the kinetic energy of the solenoid current
carriers, that the latter energy is not even mentioned.

\section{Electrostatic Analogue for the Magnetic Aharonov-Bohm Phase Shift}

\subsection{A Solenoid as a Limit of Magnetic Dipoles}

It was pointed out in 1987, that there is a classical \textit{electrostatic}
lag effect,\cite{B1987a} which is the analogue of the classical lag suggested
as the electromagnetic basis for the Aharonov-Bohm phase shift. \ A long
solenoid can be regarded as a line of magnetic dipoles per unit length.
\ Thus, each turn of radius $b$ carrying a current $I$ of the solenoid
provides a magnetic dipole $\allowbreak$of magnitude $m=\pi b^{2}I/c$, and the
number $n$ of turns per unit length converts this expression into that for a
line of magnetic dipoles. \ The magnetic dipole moment per unit length is
$\left(  \pi b^{2}I/c\right)  n=\left[  1/\left(  4\pi\right)  \right]
B_{0}\pi b^{2}$ where we have introduced the magnetic field inside a long
solenoid $B_{0}=4\pi nI/c$. \ The electrostatic analogue involves two line
charges $\pm\lambda$ per unit length of opposite sign separated by a small
perpendicular displacement $\mathbf{d}$ from $-\lambda$ to $+\lambda$ so that
the two can be described in terms of an electric dipole moment per unit length
$\mathbf{D=d}\lambda$.

\subsection{A Ring of Electric Dipoles}

In order to simplify our calculations and take advantage of axial symmetry, we
will consider a line of electric dipoles which is bent into a circular shape,
analogous to a solenoid being bent into a magnetic toroid. \ The corresponding
lines of electric charges are bent into two circular rings, each of total
charge $\pm q=\pm\left(  2\pi r_{\pm}\right)  \lambda$, one of radius
$r_{+}=R+d/2$ for the plus charge, and the second of radius $r_{-}=R-d/2$ for
the minus charge. \ 

We imagine that these rings of charge are each made up of $N$ individual
charges $\pm q_{i}$ with values $\pm q_{i}=\pm q/N$. \ The individual charges
are allowed to slide frictionlessly on straight rods parallel to the $z$-axis,
where the rods themselves are arranged in a circular pattern, with the rods at
radius $R+b/2$ carrying the positive charges $q_{i}$, and the rods at radius
$R-b/2$ carrying the negative charges $-q_{i}$. \ The charges $\pm q_{i}$ are
attracted to the charges in the ring of opposite charge, both of which,
initially, are assumed at rest in the $xy$-plane.

\subsection{Interaction of a Charge $e$ Initially at Rest and the Ring of
Electric Dipoles}

Now we introduce the charge $e$ which slides frictionlessly on a rod along the
central $z$-axis. \ When the charge $e$ is initially at rest and is located at
$z_{e}$ along the $z$-axis, there will be electrostatic forces between the
charge $e$ and the charges $\pm q_{i}$ of the rings of charge. \ However, due
to the electromagnetic inertia of the closely-spaced charges $q_{i}$ in the
plus ring, (as was discussed in the previous section), the charges $q_{i}$ of
the plus ring will have a very small common acceleration away from the
(positive) charge $e$. \ Similarly, due to the electromagnetic inertia of the
closely-space charges $-q_{i}$ in the minus ring, the charges of the minus
ring will have a very small common acceleration towards the charge $e.$ \ The
accelerations of the plus and minus rings of charge will tend to polarize
slightly the circle of electric dipoles, since the plus ring and the minus
ring have small accelerations in opposite directions.

\subsection{Interaction of a Moving Charge $e$ and the Ring of Electric
Dipoles}

We next consider an electric situation where the charge $e$ is initially
moving with a speed $v_{e0}$ along the $z$-axis when it is far away from the
rings of charge located at $z_{q}=0$. \ Since the acceleration of each ring of
charge is negligible because of its large electromagnetic inertia, the ring of
dipoles will become only very slightly polarized and will put electrostatic
dipole forces on the charge $e$ which passes through the center of the rings.
\ Thus the acceleration of the charge $e$ due to the electrostatic forces of
the ring is given by Newton's second law, which for nonrelativistic speed
$v_{e0}$ is
\begin{align}
M_{e}\frac{dv_{e}}{dt}  &  =e\left[  \frac{q(z_{e}-z_{q})}{\left[
(z_{e}-z_{q})^{2}+\left(  R+b/2\right)  ^{2}\right]  ^{3/2}}-\frac
{q(z_{e}-z_{q})}{\left[  (z_{e}-z_{q})^{2}+\left(  R-b/2\right)  ^{2}\right]
^{3/2}}\right]  _{z_{q}=0}\nonumber\\
&  =\frac{eqz_{e}}{\left(  z_{e}^{2}+R^{2}\right)  ^{3/2}}\left[  1-\frac
{3}{2}\frac{\left(  Rb+b^{2}/4\right)  }{\left(  z_{e}+R^{2}\right)
}+...-\left(  1-\frac{3}{2}\frac{\left(  -Rb+b^{2}/4\right)  }{\left(
z_{e}+R^{2}\right)  }+...\right)  \right] \nonumber\\
&  =\frac{-3eqz_{e}Rb}{\left(  z_{e}^{2}+R^{2}\right)  ^{5/2}} \label{medvt}%
\end{align}
to leading order in the ratio $b/R$. \ When the value of $z_{e}$ is negative,
the acceleration of the (positive) charge $e$ is positive.

Assuming that the interaction between the charge $e$ and the rings of electric
charge is small, we can calculate the change in velocity $\Delta v_{e}\left(
t\right)  $ of the charge $e$ at time $t$ while approximating $v_{e}%
(t)\approxeq v_{e0}$ on the right-hand side of Eq. (\ref{medvt}). \ Thus
writing $z_{e}\approxeq v_{e0}t$, we find $\Delta v_{e}\left(  t\right)
=v_{e}\left(  t\right)  -v_{e0}$ where%
\begin{equation}
\Delta v_{e}\left(  t\right)  =%
%TCIMACRO{\tint \nolimits_{-\infty}^{t^{\prime}=t}}%
%BeginExpansion
{\textstyle\int\nolimits_{-\infty}^{t^{\prime}=t}}
%EndExpansion
dt^{\prime}\left(  \frac{-3eq\left(  v_{e0}t^{\prime}\right)  Rb}{M_{e}\left[
\left(  v_{e0}t^{\prime}\right)  ^{2}+R^{2}\right]  ^{5/2}}\right)
=\frac{eqRb}{M_{e}v_{e0}\left[  \left(  v_{e0}t\right)  ^{2}+R^{2}\right]
^{3/2}}. \label{Dvett}%
\end{equation}

\subsection{Conservation of Electric Energy}

In this electrostatic case, it is easy to apply forces to calculated the
velocity change $\Delta v_{e}$ of the charge $e$. \ However, the calculations
are easy only under the assumption that the electromagnetic inertia is so
large that the rings of charge do not change their positions appreciably.
\ Under this assumption of large electromagnetic inertia for each charge ring
forming the ring of electric dipoles, we can apply energy conservation,
involving both particle kinetic energy and electric field energy. \ The charge
$e$ starts with kinetic energy $m_{e}v_{e0}^{2}/2$ when far away from the
rings of charge, and the electrostatic energy of the rings themselves is
assumed essentially unchanged during the short time that the charge $e$ is
near the dipole ring. \ Thus energy conservation for a small velocity change
of the charge $e$ requires that the change in particle kinetic energy
$M_{e}v_{e}^{2}/2-M_{e}v_{e0}^{2}/2\approxeq M_{e}v_{e0}\Delta v_{e}$ should
be balanced by the change in electric potential energy between the charge $e$
and the ring of dipoles,
\begin{align}
M_{e}v_{e0}\Delta v_{e}  &  =-e\left[  \frac{q}{\left[  (z_{e}-z_{q}%
)^{2}+\left(  R+b/2\right)  ^{2}\right]  ^{1/2}}-\frac{q}{\left[  (z_{e}%
-z_{q})^{2}+\left(  R-b/2\right)  ^{2}\right]  ^{1/2}}\right]  _{z_{q}%
=0}\nonumber\\
&  =\frac{eqRb}{\left[  z_{e}^{2}+R^{2}\right]  ^{3/2}} \label{Enmev}%
\end{align}
to lowest order in $b/R$. \ Thus the energy conservation equation
(\ref{Enmev}) gives the same velocity shift as found in Eq. (\ref{Dvett})
based on the electric forces.

\subsection{Classical Electrostatic Lag Effect}

We can integrate Eq. (\ref{Dvett}) for the change in velocity of the charge
$e$ in order to obtain a relative lag (or lead) for the charge $e$ which
passes through the ring of electric dipoles, compared to a charge which
travels with constant velocity $v_{e}\left(  t\right)  =v_{e0}$. \ The
relative lag $\Delta z_{e}$ (after the charge $e$ has passed through the ring
of electric dipoles) can be given through first order in the interaction
between the charge $e$ and the ring of electric dipoles as
\begin{align}
\Delta z_{e}  &  =%
%TCIMACRO{\tint \nolimits_{-\infty}^{\infty}}%
%BeginExpansion
{\textstyle\int\nolimits_{-\infty}^{\infty}}
%EndExpansion
dt\,\Delta v_{e}\left(  t\right)  =%
%TCIMACRO{\tint \nolimits_{-\infty}^{\infty}}%
%BeginExpansion
{\textstyle\int\nolimits_{-\infty}^{\infty}}
%EndExpansion
dt\frac{1}{M_{e}v_{e0}}\left(  \frac{eqRb}{\left(  z_{e}^{2}+R^{2}\right)
^{3/2}}\right) \nonumber\\
&  =%
%TCIMACRO{\tint \nolimits_{-\infty}^{\infty}}%
%BeginExpansion
{\textstyle\int\nolimits_{-\infty}^{\infty}}
%EndExpansion
dt\frac{1}{M_{e}v_{e0}}\left(  \frac{eqRb\,}{\left[  (v_{e0}t)^{2}%
+R^{2}\right]  ^{3/2}}\right)  =\frac{eqb}{M_{e}v_{e0}^{2}}\frac{2}{R}%
=4\pi\frac{eD}{M_{e}v_{e0}^{2}}, \label{DzE}%
\end{align}
where we have introduced the dipole moment per unit length $D=\lambda
b=\left[  q/\left(  2\pi R\right)  \right]  b$ where $b$ is the
perpendicular\ (radial) distance between the circles of charges. \ 

We notice that the final expression in Eq. (\ref{DzE}) for this
\textit{electrostatic} relative lag (between a charge with passes
\textit{through} the ring of electric dipoles and a charge $e$ which has a
constant velocity $v_{e0}$) is \textit{independent of the average radius }%
$R$\textit{ of the ring of electric dipoles}. \ The lag depends upon only
whether or not the charge $e$ has passed through the ring of electric dipoles.
\ Also, the lag is proportional to the dipole moment $D$ per unit length.
\ This electric lag situation is analogous to the \textit{magnetic} lag
situation where the lag depends upon the magnetic flux. \ In the next section
when we discuss the Aharonov-Bohm phase shift, we will obtain a similar
classical lag where the electric dipole moment per unit length $D$ for a ring
of \textit{electric} dipoles is replaced by the magnetic dipole moment per
unit length or magnetic flux $B_{0}\pi b^{2}$ of a toroid pictured as a ring
of \textit{magnetic} dipoles. \ In both cases, the classical lag is
independent of the average radius $R$ of the ring. \ If the radius $R$ is
taken ever larger, the rings of dipoles become straight lines; the lag is
relative to which side the electron passed on but is independent of the
distance between the charge $e$ and the straight line of dipoles. \ 

\section{The Aharonov-Bohm Phase Shift}

\subsection{Physical Arrangement}

The basic lag idea, which appears for the simple model above, provides a
classical electromagnetic basis for the Aharonov-Bohm interaction where a
charge $e$ interacts with a magnet. \ For simplicity of calculation, we
consider a magnetic toroid rather than a long solenoid. \ We consider a
charged particle $e$ moving along the axis of symmetry of the current-carrying
toroid. \ We will assume that the toroid has the $z$-axis as its axis of
symmetry, and, at average radius $R$, has an (average) magnetic field
\begin{equation}
\mathbf{B}_{T}=\widehat{\phi}B_{T}=\widehat{\phi}\frac{4\pi}{c}nI,
\end{equation}
where $n$ is the number of turns per unit length and $I$ is the current in
each turn. \ The magnetic field is confined to the interior of the toroidal
volume, which has a small cross-sectional area $\pi b^{2}$, and has volume
$\left(  2\pi R\right)  \pi b^{2}$. \ The toroid is electrically neutral, with
fixed negative charges balancing the moving positive charges which produce the
magnetic field $\mathbf{B}_{T}$. \ Thus, in its own rest frame, the
unperturbed toroid has neither electric nor magnetic fields outside the volume
of the toroid.\ 

\subsection{Analysis in the Inertial Frame where the Toroid is at Initially at
Rest}

\subsubsection{Magnetic Force on the Toroid Due to the Charge $e$}

We will consider the interaction of the charge $e$ and the toroid in the
lowest order approximation, where the charge $e$ and the toroid each continue
their unperturbed motions despite the presence of the other. \ In this
approximation when evaluated in the rest frame $S_{T}$ of the toroid with the
charge $e$ moving with velocity $\mathbf{v}$, the charge $e$ experiences no
forces since there are no electric or magnetic fields outside the
\textit{unperturbed} toroid. \ 

The toroid is electrically neutral and so experiences no net electric force
due to the charge $e$. \ On the other hand, the magnetic field of the moving
charge $e$ does place a net magnetic force on the toroid. \ If we consider one
turn of the toroid as a magnetic dipole
\begin{equation}
\mathbf{m}=\widehat{\phi}\pi b^{2}I/c \label{m}%
\end{equation}
where $\pi b^{2}$ is the cross-sectional area and $I$ is the current in the
turn, then the force on the turn due to the magnetic field $\mathbf{B}_{e}$ of
the charge $e$ is given by%
\begin{equation}
\mathbf{F}_{\mathbf{m}}^{(B)}=-\nabla_{\mathbf{m}}\left\{  -\mathbf{m\cdot
B}_{e}\left(  \mathbf{r}_{\mathbf{m}},t\right)  \right\}  =-\nabla
_{\mathbf{m}}\left\{  -\mathbf{m\cdot}\left[  e\frac{\mathbf{v}}{c}\times
\frac{\widehat{r}R+\widehat{z}\left(  z_{\mathbf{m}}-z_{e}\right)  }{\left[
\left(  z_{\mathbf{m}}-z_{e}\right)  ^{2}+R^{2}\right]  ^{3/2}}\right]
\right\}  . \label{FmB}%
\end{equation}
Here we can interchange the dot product and the cross product, and note that
$\mathbf{m\times v}$ is in the direction $\widehat{r}$. \ The $z$-component of
this force in Eq. (\ref{FmB}) is multiplied by the number $n2\pi R$ of rings
around the toroid, giving a force on the toroid along the axis of symmetry
\begin{equation}
F_{T}^{\left(  B\right)  }=\left[  \left(  n2\pi R\right)  \left(  \frac{\pi
b^{2}I}{c}\right)  \right]  \frac{ev}{c}\frac{-3R\left(  z_{T}-z_{e}\right)
}{\left[  \left(  z_{T}-z_{e}\right)  ^{2}+R^{2}\right]  ^{5/2}}. \label{FTB}%
\end{equation}
It appears that the forces are not balanced, since apparently only the toroid
experiences a force whereas the charge $e$ does not.

\subsubsection{Magnetic Energy Change}

The analysis in this $S_{T}$ inertial frame also raises questions of energy
conservation. \ It appears that there is no change in the kinetic energy of
the charge $e$, nor in the electric field energy during the interaction
(treated in the \textit{unperturbed} approximation), but there is a change in
the magnetic energy since the magnetic field of the moving charge $e$ overlaps
the magnetic field inside the toroid. \ Both magnetic fields are in the
$\widehat{\phi}$-direction, and the overlap energy of the magnetic fields is
\begin{equation}
\Delta U_{overlap}^{\left(  B\right)  }=\frac{1}{4\pi}\mathbf{B}_{e}%
\cdot\mathbf{B}_{T}\left(  2\pi R\pi b^{2}\right)  =\frac{1}{4\pi}\left[
e\frac{v}{c}\frac{R}{\left[  \left(  z_{\mathbf{m}}-z_{e}\right)  ^{2}%
+R^{2}\right]  ^{3/2}}\right]  \left(  \frac{4\pi}{c}nI\right)  \left(  2\pi
R\pi b^{2}\right)  . \label{DUB1}%
\end{equation}
However, the balancing aspect for this magnetic energy change is not
immediately apparent.

\subsection{Analysis in the Inertial Frame where the Charge $e$ is Initially
at Rest}

\subsubsection{Change of Inertial Frame Used in the Description}

For a relativistic system, a change of inertial frames may significantly
change the description of the physical system. \ Thus, if we go to the
$S_{e}^{\prime}$ inertial frame where the unperturbed charge $e$ is at rest
and the unperturbed toroid is moving with velocity $-\mathbf{v}$, then there
are no \textit{magnetic} forces between the charge $e$ and the toroid, but
there are \textit{electric} forces. \ Since the charge $e$ is at rest, it has
only an electro\textit{static} field. \ In this $S_{e}^{\prime}$ inertial
frame in which it is moving with velocity $-\mathbf{v}$, the toroid now has a
circle of electric dipoles which will both generate electric fields and also
experience electric forces. \ The \textit{electric} forces between the charge
$e$ and the toroid are now equal and opposite. \ Also, in this $S_{e}^{\prime
}$ inertial frame, there is a change in energy in the electric field, and the
work done by the electric forces to change the kinetic energy of the toroid
accounts for the change in electric field energy. \ We will go through all
this analysis in detail.

\subsubsection{Lorentz Transformation for a Magnetic Dipole}

If we take one turn of the toroid (one current loop), we can picture it as a
magnetic dipole $\mathbf{m}$ as indicated in Eq. (\ref{m}). \ We wish to
consider the appearance of this magnetic dipole in an inertial frame in which
it is moving. \ The magnetic dipole $\mathbf{m}$ can be pictured in its rest
frame as a circle in the $zx$-plane of radius $b$ with equally-spaced charges
with the ring's center at the origin. \ The negative charges $-q_{i}$ are
stationary at coordinates $z_{-i}=b\cos\phi_{i}$, $x_{-i}=b\sin\phi_{i}$; the
moving positive charges are located at $z_{+i}=b\cos\left(  \omega t+\phi
_{i}\right)  $, $x_{+i}=b\sin\left(  \omega t+\phi_{i}\right)  $ while moving
in a circular orbit\ with speed $v=\omega b.$ \ We are interested in the
trajectories of these charges as seen in a primed inertial frame moving with
velocity $\mathbf{v=}\widehat{z}v$ along the $z$-axis, so that the magnetic
dipole now has a velocity $-\mathbf{v}$. \ 

For the small speeds $v<<c$ \ which are appropriate for the Darwin Lagrangian,
the Lorentz transformations involve $z=\gamma\left(  z^{\prime}+vt^{\prime
}\right)  \approxeq\left[  1+v^{2}/(2c^{2})\right]  \left(  z^{\prime
}+vt^{\prime}\right)  $, $x=x^{\prime}$, and $t=\gamma\left(  t^{\prime
}+vz^{\prime}/c^{2}\right)  \approxeq\left[  1+v^{2}/(2c^{2})\right]  \left(
t^{\prime}+vz^{\prime}/c^{2}\right)  \approxeq\left[  1+v^{2}/(2c^{2})\right]
t^{\prime}+vz^{\prime}/c^{2}$. \ Then as seen in the primed frame where the
magnetic dipole is moving with velocity $-\mathbf{v}$, the locations of the
negative charges are at primed coordinates
\begin{equation}
\left(  z_{-i}^{\prime}+vt^{\prime}\right)  =+b\left[  1-v^{2}/(2c^{2}%
)\right]  \cos\phi_{i},\text{ \ \ and \ \ }x_{-i}^{\prime}=b\sin\phi_{i},
\label{negppi}%
\end{equation}
corresponding to a contraction in the direction of motion. \ 

The trajectories of the current-carrying charges are found from expanding
\begin{align}
\left(  z_{+i}^{\prime}+vt^{\prime}\right)   &  =\frac{1}{\gamma}z_{+i}%
=\frac{b}{\gamma}\cos\left\{  \omega\gamma\left(  t^{\prime}+vz_{+i}^{\prime
}/c^{2}\right)  +\phi_{i}\right\} \nonumber\\
&  \approxeq\left[  1-v^{2}/(2c^{2})\right]  b\cos\left\{  \omega\left[
1+v^{2}/(2c^{2})\right]  t^{\prime}+\phi_{i}+\omega vz_{+i}^{\prime}%
/c^{2}\right\}
\end{align}
and
\begin{align}
x_{+i}^{\prime}  &  =x_{+i}=b\sin\left\{  \omega\gamma\left(  t^{\prime
}+vz_{+i}^{\prime}/c^{2}\right)  +\phi_{i}\right\} \nonumber\\
&  \approxeq b\sin\left\{  \omega\left[  1+v^{2}/(2c^{2})\right]  t^{\prime
}+\phi_{i}+\omega vz_{+i}^{\prime}/c^{2}\right\}  .
\end{align}
Expanding these expressions for small speeds, $v<<c$, we find
\begin{align}
\left(  z_{+i}^{\prime}+vt^{\prime}\right)   &  =\left[  1-v^{2}%
/(2c^{2})\right]  b\cos\left\{  \omega\left[  1+v^{2}/(2c^{2})\right]
t^{\prime}+\phi_{i}\right\} \nonumber\\
&  -b\frac{\omega v}{c^{2}}\frac{b}{2}\sin\left\{  2\left(  \omega\left[
1+v^{2}/(2c^{2})\right]  t^{\prime}+\phi_{i}\right)  \right\}  \label{zppi}%
\end{align}
and%
\begin{equation}
x_{+i}^{\prime}=b\sin\left\{  \omega\left[  1+v^{2}/(2c^{2})\right]
t^{\prime}+\phi_{i}\right\}  +b\frac{\omega v}{c^{2}}b\cos^{2}\left\{
\omega\left[  1+v^{2}/(2c^{2})\right]  t^{\prime}+\phi_{i}\right\}  .
\label{xppi}%
\end{equation}
When treated in the small-source approximation, the averages over the phases
$\phi_{i}$ at time $t^{\prime}$ (for both the positive charges and negative
charges) agree on the average $z$-component location, $\left\langle
z_{+i}^{\prime}\right\rangle =$ $-vt^{\prime}=\left\langle z_{-i}^{\prime
}\right\rangle $. \ However, the moving positive charges now have an electric
dipole moment (perpendicular to the velocity direction), since when averaged
over the phases $\phi_{i}$, we find from Eq. (\ref{xppi}) that%
\begin{equation}
q\left\langle x_{+i}^{\prime}\right\rangle =\frac{v}{c}\frac{q\omega b^{2}%
}{2c}=\frac{v}{c}m \label{dip}%
\end{equation}
where $m$ is the magnetic dipole moment of the arrangement, while
$\left\langle x_{-i}^{\prime}\right\rangle =0$. \ Our calculation is entirely
in agreement with the first edition\cite{Jackson1} of Jackson's text on
classical electrodynamics and with a problem in Griffiths\cite{G572}, where it
is pointed out that a magnetic dipole moving with velocity $\mathbf{v}$ has an
electric dipole in lowest order given by $\mathbf{p}=(\mathbf{v}%
/c)\times\mathbf{m}$. \ In our case, the direction if the dipole moment is
given by $\left(  -\widehat{z}\right)  \times\widehat{\phi}=\widehat{r}$,
where $\widehat{r}$ points from the $z$ axis out to the location of the
magnetic dipole in the $xz$-plane.

In the $S_{e}^{\prime}$ inertial frame, the electric field $\mathbf{E}%
^{\prime}\left(  \mathbf{r}^{\prime},t^{\prime}\right)  $ for a field point
$\mathbf{r}^{\prime}$ in the $x^{\prime}z^{\prime}$-plane can be found by
substituting the components $z_{-i}^{\prime}$, $x_{-i}^{\prime}$,
$z_{+i}^{\prime}$, and $x_{+i}^{\prime}$ in Eqs. (\ref{negppi}), (\ref{zppi})
and (\ref{xppi}) into Eq. (\ref{Ert}). \ We note that the expression in Eq.
(\ref{Ert}) involves an electrostatic contribution plus additional terms in
$1/c^{2}.$ \ Since terms in $1/c^{4}$ are neglected, the only place where the
$1/c^{2}$ terms in Eqs. (\ref{negppi}), (\ref{zppi})\ and (\ref{xppi}) will
actually contribute to the electric field in order $1/c^{2}$ is through the
electrostatic contribution $\mathbf{E}^{\prime}\left(  \mathbf{r}^{\prime
},t^{\prime}\right)  =%
%TCIMACRO{\tsum \nolimits_{a}}%
%BeginExpansion
{\textstyle\sum\nolimits_{a}}
%EndExpansion
\left(  \mathbf{r}^{\prime}-\mathbf{r}_{a}^{\prime}\right)  /\left\vert
\mathbf{r}^{\prime}-\mathbf{r}_{a}^{\prime}\right\vert ^{3}$. \ Thus the
electric dipole moment found in Eq. (\ref{dip}) will contribute in a form
analogous to that familiar from electrostatics.

\subsubsection{Electric Forces}

The toroid of our analysis involves not a single current loop forming a
magnetic dipole but rather an array of such magnetic dipoles arranged in a
circular pattern around the axis of symmetry (our $z$-axis) of the toroid.
\ Thus, in the $S_{e}^{\prime}$ inertial frame in which the charge $e$ is at
rest and the toroid is moving with velocity $-\mathbf{v}$, the toroid exhibits
a circular array of electric dipoles which is exactly analogous to the ring of
electric dipoles discussed above in the section involving the electrostatic
analogy. \ However, now the electric dipoles are of order $1/c^{2}$, not
zero-order in $1/c$. It is these electric dipoles of order $1/c^{2}$ which
produce an \textit{electric} force on the charge $e$ and experience an
\textit{electric} force on the toroid. \ 

The net force on the toroid corresponds to the sum of the $z$-components of
the force on each of the electric dipoles and is given by%
\begin{align}
F_{T}^{\left(  E\right)  }  &  =%
%TCIMACRO{\tsum \nolimits_{i}}%
%BeginExpansion
{\textstyle\sum\nolimits_{i}}
%EndExpansion
\widehat{z}\cdot\left[  (\mathbf{p}_{i}\mathbf{\cdot}\nabla_{\mathbf{m}%
})\mathbf{E}_{e}\left(  \mathbf{r}_{m},t\right)  \right] \nonumber\\
&  =\left(  n2\pi R\right)  \left(  \frac{v}{c}m\right)  \frac{\partial
}{\partial R}e\frac{(z_{T}-z_{e})}{\left[  \left(  z_{T}-z_{e}\right)
^{2}+R^{2}\right]  ^{3/2}}\nonumber\\
&  =\left(  n2\pi R\right)  \frac{v}{c}\left(  \frac{\pi b^{2}I}{c}\right)
e\frac{-3R(z_{T}-z_{e})}{\left[  \left(  z_{T}-z_{e}\right)  ^{2}%
+R^{2}\right]  ^{5/2}}. \label{FTE}%
\end{align}
This electric force in Eq. (\ref{FTE}) is the same as the magnetic force which
appears in Eq. (\ref{FTB}). \ In that equation (\ref{FTB}), the force is a
\textit{magnetic} force on the toroid appearing in the $S_{T}$ inertial frame
in which the toroid is at rest and the charge $e$ is moving. \ In equation
(\ref{FTE}), the force is an \textit{electric} force appearing in the
$S_{e}^{\prime}$ inertial frame where the charge $e$ is at rest and the toroid
is moving. \ Since our analysis, like the Darwin Lagrangian, is accurate
through order $1/c^{2}$, we expect these forces to be unchanged in magnitude
on change of inertial frames for speeds $v<<c$. \ The electric force
$F_{T}^{\left(  E\right)  }$ of the charged particle on the toroid has its
electric partner in the force $F_{e}^{\left(  E\right)  }$ of the toroid on
the charge $e,$ $F_{e}^{\left(  E\right)  }=-F_{T}^{\left(  E\right)  },$
which force is equal in magnitude and opposite in direction, due to the circle
of electric dipoles appearing in the toroid. \ 

\subsubsection{Electric Energy}

In the $S_{e}^{\prime}$ inertial frame, the change in the energy in the
electric fields is associated with the energy of an electric dipole in the
field of a point charge%
\begin{equation}
\Delta U^{\left(  E\right)  }=-%
%TCIMACRO{\tsum \nolimits_{i}}%
%BeginExpansion
{\textstyle\sum\nolimits_{i}}
%EndExpansion
\mathbf{p}_{i}\mathbf{\cdot E}=-\left(  n2\pi R\right)  \left[  \left(
\frac{v}{c}\right)  \left(  \frac{\pi b^{2}I}{c}\right)  \right]  e\frac
{R}{\left[  \left(  z_{T}-z_{e}\right)  ^{2}+R^{2}\right]  ^{3/2}}.
\label{DUE1}%
\end{equation}
The work done by the electric forces in Eq. (\ref{FTE}) is connected directly
to changes in electrical energy in the electromagnetic fields.\cite{Forces}
\ It is easy to check that $\Delta U^{\left(  E\right)  }=-%
%TCIMACRO{\tint \nolimits_{\infty}^{z_{T}}}%
%BeginExpansion
{\textstyle\int\nolimits_{\infty}^{z_{T}}}
%EndExpansion
dzF_{T}^{\left(  E\right)  }$. Thus in the $S_{e}^{\prime}$\ inertial frame in
which the charge $e$ is at rest and the toroid is moving, we have balance
between the work done and the changes in electric energy. \ However, we see
that the change in electric energy $\Delta U^{\left(  E\right)  }$ in Eq.
(\ref{DUE1}) is just the \textit{negative} of the positive magnetic overlap
energy $\Delta U_{overlap}^{\left(  B\right)  }$ in Eq. (\ref{DUB1}). \ 

\subsection{Reanalysis of the Charge-Toroid Interaction in the Toroid Inertial
Frame}

\subsubsection{Two Aspects of Magnetic Energy Change}

The interaction of the unperturbed motions of the charge $e$ and toroid seems
both transparent and energy-balanced in the $S_{e}^{\prime}$ frame where the
charge $e$ is at rest and the toroid is moving with velocity $-\mathbf{v}$.
\ Why does the analysis seem so different in the $S_{T}$\ rest frame of the
toroid where the charge $e$ is moving with velocity $\mathbf{v}$? \ Part of
the answer lies in the fact that an aspect of energy transfer between the
charge $e$ and the toroid has not yet been mentioned. \ 

The magnetic energy change of the entire system\ here has two terms and takes
the form%

\begin{equation}
\Delta U^{\left(  B\right)  }=\Delta U_{overlap}^{\left(  B\right)  }+\Delta
U_{toroid\text{ }currents}^{\left(  B\right)  }\text{.}%
\end{equation}
Thus, in addition to the term $\Delta U_{overlap}^{\left(  B\right)  }$, there
is a second term involving transfer of energy as the electric fields of the
charge $e$ act on the currents of the toroid. \ It is not the
electro\textit{static} field of the charge $e$ (which has no $emf$) which
transfers net energy to the toroid currents, but rather the terms in $1/c^{2}$
in the electric field of the charge $e$. \ These $1/c^{2}$ electric field
terms, which are consistent with the changing magnetic field $\mathbf{B}_{e}$,
cause changes in the speed $\Delta v_{i}$ of the toroid particles $q_{i}$ and
hence changes in the toroid magnetic field $\mathbf{B}_{T}$. \ 

\subsubsection{Response of the Toroid to Changing Magnetic Flux}

The toroid responds in the fashion of a solenoid to the changing flux due to
the moving charge $e$. \ If there is no resistance and the magnetic energy
dominates the kinetic energy of the current carriers, then \textit{the
magnetic flux in the toroid does not change}. \ Dividing out the constant
cross-sectional area $\pi b^{2}$ of the toroid, the requirement of constant
magnetic flux for the toroid gives%
\begin{equation}
0=\Delta B_{e}+\Delta B_{T}=B_{e}+\left(  B_{T}-B_{T0}\right)  ,
\end{equation}
since initially the magnetic field $B_{e}$ of the charge $e$ at the position
of the toroid vanishes, while the toroid is carrying currents giving an
initial magnetic field $B_{T0}$. Thus constant flux for the toroid means that
we have
\begin{equation}
B_{T}-B_{T0}=-B_{e}.
\end{equation}

\subsubsection{Magnetic Energy Change}

The magnetic energy in the overlap integral is linear in the interaction and
involves sums of the form $ev_{e}%
%TCIMACRO{\tsum \nolimits_{i}}%
%BeginExpansion
{\textstyle\sum\nolimits_{i}}
%EndExpansion
q_{i}v_{i}$. \ On the other hand, the change in the magnetic energy of the
toroid is quadratic in the interaction and involves sums of the form $%
%TCIMACRO{\tsum \nolimits_{i}}%
%BeginExpansion
{\textstyle\sum\nolimits_{i}}
%EndExpansion%
%TCIMACRO{\tsum \nolimits_{j\neq i}}%
%BeginExpansion
{\textstyle\sum\nolimits_{j\neq i}}
%EndExpansion
q_{i}q_{j}\left[  (v_{i}+\Delta v)\left(  v_{j}+\Delta v\right)  -v_{i}%
v_{j}\right]  =%
%TCIMACRO{\tsum \nolimits_{i}}%
%BeginExpansion
{\textstyle\sum\nolimits_{i}}
%EndExpansion%
%TCIMACRO{\tsum \nolimits_{j\neq i}}%
%BeginExpansion
{\textstyle\sum\nolimits_{j\neq i}}
%EndExpansion
q_{i}q_{j}[v_{i}\Delta v+v_{j}\Delta v]$. \ Then keeping the lowest order
terms, the total change of magnetic energy $\Delta U^{\left(  B\right)  }$ in
the toroid involves a constant volume $\pi b^{2}2\pi R$ of the toroid
multiplying the changes in magnetic energy density%
\begin{align}
\Delta U^{\left(  B\right)  }  &  =\frac{1}{4\pi}\left[  B_{e}B_{T0}%
-2B_{e}B_{T0}\right]  \left(  \pi b^{2}2\pi R\right) \nonumber\\
&  =\frac{1}{4\pi}\left[  -B_{e}B_{T0}\right]  \left(  \pi b^{2}2\pi R\right)
=-\Delta U_{overlap}^{\left(  B\right)  }.
\end{align}
Thus, the \textit{total} change in magnetic field energy is exactly the
\textit{negative} of the \textit{overlap} magnetic energy $\Delta
U_{overlap}^{\left(  B\right)  }$ appearing in Eq. (\ref{DUB1}). \ 

\subsubsection{Agreement with the Electric Energy Change}

The total magnetic energy change \ $\Delta U^{\left(  B\right)  }$ agrees
exactly with the electric energy change found in the inertial frame
$S_{e}^{\prime}$ and appearing in Eq. (\ref{DUE1}). \ The $1/c^{2}$
electromagnetic energy changes for the system (of the charge $e$ and the
toroid) are the same in both inertial frames. \ However, in the $S_{T}$ frame,
the electromagnetic energy change is magnetic $\Delta U^{\left(  B\right)  }$,
while in the $S_{e}^{\prime}$ inertial frame, the electromagnetic energy
change is electric $\Delta U^{\left(  E\right)  }$. \ The magnetic energy
change is accounted for by the work done by the back (Faraday) acceleration
electric fields which keep the magnetic flux in the toroid constant and also
put a magnetic-energy-balancing force back on the charge $e$ which is the
agent trying to change the magnetic field in the toroid.\cite{Forces} \ It is
the speed changes of the toroid current carriers which produce a back
(Faraday) electric field on the charge $e$ in the inertial frame $S_{T}$ in
which the toroid is at rest. \ 

\subsubsection{Failure of the Hierarchy of Powers of $1/c$}

In the past, it has been claimed that the electric fields of the moving charge
$e$ cannot possibly provide a force back on the passing charge which is of
order $1/c^{2}$. \ It was claimed that the induction fields of the charge $e$
will cause the current-carrying charges of the toroid to accelerate in order
$1/c^{2}$ and hence produce back (Faraday) acceleration electric fields at the
charge $e$ of order $1/c^{4}$. \ This argument involving powers in $1/c$ is
fallacious. \ It is false because it does not correspond to how a toroid (or
solenoid) responds to an agent trying to change the speed of its current
carriers. \ A solenoid with a large value of inductance $L$ will try to
prevent a change in the speed of its current carriers and so place a back
force on the agent trying to cause the speed change. \ And because there are a
vast number of current carriers in the solenoid, the back force does not
observe the hierarchy of orders in $1/c$. \ For example, the charge carriers
of a solenoid have masses which are zero-order in $1/c$; the back force due to
the back (Faraday) acceleration field of each current carrier may be of order
$1/c^{2}$, but the vast number of current carriers produce a back electric
force which tries to stops the agent from changing the speed of the current
carriers because a change in speed will lead to a change in magnetic energy.
\ It is this same back (Faraday) acceleration electric field which acts on the
passing electron to change its kinetic energy so as to achieve magnetic energy
balance for the entire system of the charge $e$ and the toroid. \ 

\subsubsection{Unfamiliarity of the Acceleration Electric Fields in
Quasistatic Electrodynamics}

In recent textbooks, the crucial (Faraday) acceleration terms appearing in the
electric field of Eq. (\ref{Ert}) are never mentioned. \ The textbook
calculations of $emfs$ are made for symmetric situations where it is easy to
find the electric field from changing magnetic fluxes without ever mentioning
the acceleration electric fields of charges. \ Indeed, usually the charge
carriers in the solenoid are never mentioned at all, since their kinetic
energy is miniscule compared to the energy in the magnetic fields of the
solenoid. \ To be sure, the acceleration electric fields did appear in a
textbook\cite{PA} in 1940, but they have disappeared in recent treatments. \ 

Unfamiliarity with the acceleration fields which balance changes in magnetic
energy has led to a whole class of \textquotedblleft
paradoxes\textquotedblright\ involving the interaction of charged particles
and magnets. \ Whenever the speeds of charged particles are changed, there
will be changes in magnetic field energy. \ These energy changes are not
accounted for by magnetic forces, which do no work. \ \textit{Magnetic energy
balance for quasistatic systems requires the existence of forces associated
with the accelerations of charged particles.}\cite{Forces} \ The changes in
the speeds of charged particles leads to acceleration fields, as given here in
the last line of Eq. (\ref{Ert}). \ These acceleration fields place a force on
the agent (which is producing the changes in the speeds of the charged
particles) in an effort to enforce energy balance (Lenz's law). \ 

\subsection{Classical Lag for the Charge $e$ in the Aharonov-Bohm Situation}

\subsubsection{Velocity Change for the Charge $e$}

The energy balance in the $S_{T}$ rest frame of the toroid is suppled by the
change in velocity of the charge $e$ which compensates for the change in total
magnetic energy given as the negative of Eq. (\ref{DUB1}),%
\begin{align}
0  &  =M_{e}v_{e0}\Delta v_{e}+\Delta U^{\left(  B\right)  }\nonumber\\
&  =M_{e}v_{e0}\Delta v_{e}-\frac{1}{4\pi}\left[  e\frac{v}{c}\frac{R}{\left[
\left(  z_{\mathbf{m}}-z_{e}\right)  ^{2}+R^{2}\right]  ^{3/2}}\right]
\left(  \frac{4\pi}{c}nI\right)  \left(  2\pi R\pi b^{2}\right)  \mathfrak{,}%
\end{align}
where the quantity $B_{T}=\left(  4\pi/c\right)  nI$ is the magnetic field of
the toroid which is in the same $\widehat{\phi}$-direction as the magnetic
field of the charge $e$. \ \ Then dividing out the common factor of $v_{e0}%
=v$, we find%
\begin{equation}
M_{e}\Delta v_{e}=\frac{e}{2c^{2}}\frac{B_{T}\pi b^{2}R^{2}}{\left\vert
\left(  z_{e}-z_{q}\right)  ^{2}+R^{2}\right\vert ^{3/2}}. \label{MeDv}%
\end{equation}

\subsubsection{Classical Electromagnetic Lag}

From the change in speed for the passing charge $e$ appearing in Eq.
(\ref{MeDv}), we can calculate the classical electromagnetic lag of the charge
$e$ compared to a particle with the same initial speed $v_{e0}$ which did not
pass through the magnetic toroid. \ We integrate the change in speed $\Delta
v_{e}$ over all time (just as in Eq. (\ref{DzE}) for the electrostatic
analogue) giving%
\begin{equation}
\Delta z_{e}=%
%TCIMACRO{\tint \nolimits_{-\infty}^{\infty}}%
%BeginExpansion
{\textstyle\int\nolimits_{-\infty}^{\infty}}
%EndExpansion
dt\left(  \frac{e}{2M_{e}c}\frac{B_{0}\pi b^{2}R^{2}}{\left\vert \left(
z_{e}-z_{q}\right)  ^{2}+R^{2}\right\vert ^{3/2}}\right)  =\frac{eB_{0}\pi
b^{2}}{M_{e}v_{e0}c}. \label{lag}%
\end{equation}
The magnitude of this expression for the classical electromagnetic lag is the
same as that given several times earlier,\cite{B1973c}\cite{B1987a} and leads
to exactly the interference pattern shift observed experimentally.\cite{M1962}%
\cite{Bayh1962} \ From the present analysis, one can obtain the definite
classical prediction for the direction of deflection. \ Assuming a positive
charge, and the magnetic field of the charge aligned with the magnetic field
of the toroid, the passing charge going through the center of the toroid is
first accelerated and then slowed so that $\Delta z_{e}$ is a positive
displacement. \ 

\subsubsection{Phase Shift Arising from a Classical Lag}

The Aharonov-Bohm phase shift is perhaps clearest in the arrangement of two
small openings in a mask.\cite{B1987a} \ The traditional arrangement places a
long solenoid between the two openings. \ In our case involving a toroid, one
could place the toroid around just one of the two openings in the mask.
\ Alternatively, one could take two toroids, each with half the flux; one
forward-facing toroid is placed around one opening and the other
backward-facing toroid is placed around the other opening. The shift in the
double-opening interference pattern would depend upon the current in the
toroid(s) which produces a relative lag (or lead, depending on the direction
of the current) compared to particles passing through the other opening in the
mask. \ \ 

\subsubsection{Phase Shift in Quantum Theory}

The quantum analysis for the electron interference pattern involves a shift in
the phase for the wave function $\psi(z,t)\approxeq\exp\left[  i\left(
p_{z}z/\hbar\right)  +i\phi-i\omega t\right]  $ associated with the opening
through which the electron passed. \ The Schroedinger equation indeed suggests
a phase change based upon the magnetic flux in the toroid, but does not
indicate the physical basis for the phase shift; the Schroedinger equation
does not indicate whether the phase shift is due to a velocity-change
classical lag or is due to a new constant-particle-velocity quantum
topological effect. \ At present, the experimental results allow either
interpretation.\ \ In the lag interpretation suggested in this article, the
phase difference $\phi$ between the two paths follows from Eq. (\ref{lag}) and
\ is given by
\begin{equation}
\phi=\frac{p_{z}\Delta z_{e}}{\hbar}=\left(  M_{e}v\right)  \left(
\frac{eB_{T}\pi b^{2}}{M_{e}v_{e0}c}\right)  \frac{1}{\hbar}=\frac{e\Phi
}{c\hbar}%
\end{equation}
where $\Phi=B_{T}\pi b^{2}$ is the magnetic flux in the toroid. \ The
magnitude of this phase shift is the same as predicted by Aharonov and Bohm.

\subsubsection{Both the Classical Lag and the Quantum Phase Shift Give a
Deflection Angle Independent of $\hbar$}

If the separation between the two openings in the mask corresponds to a
distance $d$, the angle of deflection\cite{angle} of the double-slit
interference as the charge $e$ passed to a distant screen would be%
\begin{equation}
\theta\approxeq\frac{\Delta z_{e}}{d}=\frac{eB_{0}\pi b^{2}}{M_{e}v_{e0}%
cd}=\frac{e\Phi}{cpd}.
\end{equation}
This deflection angle is simply a ratio of two lengths and has no dependence
on Planck's constant $\hbar$. \ The magnitude of the predicted deflection
angle $\theta$ is the same in both classical theory and in quantum theory. The
angle of deflection of the double-slit interference pattern would take the
same form for a wave of any wavelength, and would be observed for light which
passed through a piece of glass behind only one hole of the two-hole mask. \ 

\subsubsection{Direction of the Deflection}

Although the magnitude of the deflection angle $\theta$ is the same in
classical and quantum theory, the predictions for the direction of deflection
may not be the same. \ Attempts to use lag effects involving merely the
overlap magnetic energy change between the toroid and the passing charge lead
to a prediction of a lag, and hence a deflection angle, in the opposite
direction from that given here. \ Conversations with the few physicists
willing to suggest a direction of deflection from quantum theory suggest a
quantum deflection direction opposite from that required by the present
classical electromagnetic analysis.\cite{direction} \ 

\section{Discussion of Electrodynamic Interactions of Charges and Magnets}

\subsection{Problems of Charges and Magnets}

The Aharonov-Bohm phase shift is one example of a class of problems involving
the interactions of charges and magnets which have long troubled physicists.
\ The class includes, in addition to the Aharonov-Bohm phase
shift,\cite{AB1959} the Aharonov-Casher phase shift,\cite{AC} the
Shockley-James paradox,\cite{SJ} the idea of \textquotedblleft hidden
mechanical momentum in magnets,\textquotedblright\cite{hid}\ and Mansuripur's
erroneous claim\cite{Mans} of the inconsistency of special relativity.

\subsection{Vector Potential Along the Symmetry Axis of the Toroid}

Solenoids and toroids make for interesting classical electromagnetic systems
because the magnetic fields of the current carriers combine in such a way as
to tend to confine the magnetic fields to the interiors of the systems.
\ However, even though the magnetostatic fields are confined to the interiors,
changing currents lead to electromagnetic fields outside the solenoids and
toroids. \ 

Indeed, it seems interesting to note that for the toroid considered above, the
vector potential $\mathbf{A}$ can be obtained easily along the axis of the
toroid by analogy with the magnetic field along the axis of a circular current
loop of radius $R$ (exploiting the analogy $\nabla\times\mathbf{A=B,~}%
\nabla\cdot\mathbf{B=0}$ and $\nabla\times\mathbf{B=}\left(  4\pi/c\right)
\mathbf{J,}$ $\nabla\cdot\mathbf{A=0}$) as%
\begin{equation}
\mathbf{A}_{T}\mathbf{(}z\mathbf{)}=\widehat{z}\frac{c}{4\pi}\frac{2\pi
R^{2}(B_{T}\pi b^{2})}{c\left(  z^{2}+R^{2}\right)  ^{3/2}}\mathbf{.}%
\end{equation}
Magnetic energy calculations above in the inertial frame where the toroid is
at rest and the charge $e$ is moving with velocity $\mathbf{v}$ can be made
using this magnetic vector potential. \ On the other hand, in the inertial
frame where the charged particle is at rest and the toroid has a ring of
electric dipole moments $-\left(  \mathbf{v/}c\right)  \times\mathbf{m}$, the
electric energy calculations can be made using the electrostatic potential.

\subsection{Mutual Inductance of the Toroid and a Circuit Along the Symmetry
Axis \ }

We can consider coupling the toroid with a circuit which places a wire down
the symmetry axis of the toroid and closes the circuit at a great distance
from the toroid. \ Although the toroid has no magnetic fields outside its
winding for static currents, there is a mutual inductance between the toroid
and the line-current circuit given by
\begin{equation}
M=\frac{4\pi}{c^{2}}n\pi b^{2}.
\end{equation}
We notice that this mutual inductance does not depend upon the average radius
$R$ of the toroid, but does depend upon the number of turns per unit length
$n$ and the cross-sectional area $\pi b^{2}$. \ These are the factors related
to the magnetic flux in the toroid. \ If we integrate the vector potential
along the axis of symmetry of the toroid (the $z$-axis), we find%
\begin{equation}%
%TCIMACRO{\tint \nolimits_{-\infty}^{\infty}}%
%BeginExpansion
{\textstyle\int\nolimits_{-\infty}^{\infty}}
%EndExpansion
dz\mathbf{A(}z\mathbf{)=}%
%TCIMACRO{\tint \nolimits_{-\infty}^{\infty}}%
%BeginExpansion
{\textstyle\int\nolimits_{-\infty}^{\infty}}
%EndExpansion
dz\frac{c}{4\pi}\frac{2\pi R^{2}(B_{T}\pi b^{2})}{c\left(  z^{2}+R^{2}\right)
^{3/2}}=B_{T}\pi b^{2},
\end{equation}
which is the magnetic flux in the toroid. \ \ 

\subsection{Localized Pulses versus Uniform Current}

Although mutual inductance between circuits is familiar in classical
electromagnetism, the mutual-inductance approximation assumes that the current
$I$ is \textit{uniform} throughout each circuit. \ In contrast with this
situation, it is the \textit{localized-transient-pulse} aspect which is so
unfamiliar in the classical analysis of the interaction of a moving charge $e$
and a toroid (or solenoid). \ The exotic nature of the interaction lends
itself to exotic claims.

\section{Some Historical Comments}

\subsection{The Aharonov-Bohm Phase Shift}

In 1959, Aharonov and Bohm\cite{AB1959} suggested that the electromagnetic
potentials played a new role in quantum theory as compared to their role in
classical electrodynamics. \ In quantum theory, the potentials took on a new
reality, in contrast to their subordinate role in classical physics as mere
supporting auxiliaries to the force-producing fields $\mathbf{E}$ and
$\mathbf{B}$. The suggestion that the interference pattern involving electrons
would be altered by magnetic fluxes was tested experimentally with a magnetic
whisker by Chambers\cite{C1960} in 1960, and then with a current-carrying
solenoid by Moellenstedt and Bayh\cite{M1962}\cite{Bayh1962} in 1962.
\ Moellenstedt and coworkers changed the current in a microsolenoid and showed
that the double-slit interference pattern of the passing electrons moved
continuously with the current of the solenoid. \ The shift of the electron
interference pattern associated with the change in the flux in the solenoid
was regarded as surprising since a long unperturbed solenoid has no magnetic
field outside the solenoid. \ 

The new experimental results, along with the interpretation provided by
Aharonov and Bohm, was enthusiastically embraced by Feynman who wrote of the
situation with characteristic clarity in the \textit{Feynman Lectures}%
.\cite{F1964} \ Unfortunately, Feynman misinterpreted both the phase shift and
a proposed classical analogue, since he indicated that the entire electron
interference pattern was shifted rather than simply the double-slit electron
pattern within the unchanged single-slit envelope. \ When the errors were
called to his attention, Feynman agreed that he had made an error, which, had
he been aware, he would have corrected when the \textit{Lectures} were being
prepared. \ The \textit{Feynman Lectures} are still republished without any
changes and are now available to read free of charge on the internet.
\ However, Vol. II, Section 15.5 still contains the same errors.

The literature regarding the Aharonov-Bohm phase shift is vast. \ Review
articles presenting the mainstream point of view are given by Olarius and
Popescu\cite{OP} and by Batelaan and Tonomura.\cite{BT} \ Most
textbooks\cite{Gq}\cite{Ball} of quantum theory now discuss the effect. \ 

\subsection{Dissent from the Mainstream Interpretation}

Although the interpretation given by Aharonov and Bohm permeates the
mainstream physics literature today, there has been a very low undercurrent of
dissent. \ Benjamin Liebowitz\cite{Lieb} suggested in 1965 that the observed
phase shift might be due to a lag effect introduced by new
(nonelectromagnetic) forces. \ There have been reanalyses of the classical
electrodynamic aspects of the interaction of a point charge with a
solenoid.\cite{Trammel} \ Also, it was pointed out\cite{B1973b} that if the
currents of the solenoid were held constant, then the electromagnetic energy
and electromagnetic momentum changed as a charged particle passed the solenoid
since the electromagnetic fields of the charge and the solenoid overlapped.
\ If one balanced these changes in electromagnetic energy and electromagnetic
momentum with changes in the mechanical energy and momentum of the passing
electron, the magnitude of the observed phase shift could be accounted for in
terms of a classical lag effect.\cite{B1973c} \ Thus perhaps the observed
phase shift could be understood as based upon a classical electrodynamic
interaction. \ To counter this argument, a conducting sleeve was placed around
a microsolenoid, in order to screen out the electromagnetic fields of the
passing charge. \ The magnetic phase shift persisted despite the conducting
shield. \ However, it was pointed out that the screening of magnetic
\textit{velocity} fields was completely different from that for
electromagnetic \textit{waves},\cite{B1996}\cite{J337} and so the presence of
a conducting sleeve around the solenoid did not rule out the lag-based
explanation.\cite{B1999}\cite{B2000}

As interest in the phase shift continued, it was pointed out that an
electrostatic analogue of the energy shifts of the magnetic situation was
provided when the solenoid (a line of magnetic dipoles) was replaced by a line
of electric dipoles.\cite{B1987a} \ In the electrostatic case, there were
clear electrostatic forces giving rise to a classical lag effect which took
the same form as the observed magnetic phase shift. The phase shift when
electrons passed around the line of electric dipoles was confirmed
experimentally by Matteucci and Pozzi.\cite{MP}

\subsection{Optical Analogy with the Observed Phase Shift}

It was emphasized that the classical-lag interpretation of the magnetic
Aharonov-Bohm phase shift was analogous to the optical phase shift occurring
when a piece of glass was placed behind only one slit of a double-slit mask
producing a double-slit interference pattern.\cite{B1987a} \ The slowing of
the electromagnetic waves while passing through the piece of glass would lead
to a lag, and hence to a deflection of the double-slit interference pattern,
while leaving the single-slit envelope undisplaced, in exact analogy with the
observed magnetic phase shift for electrons. \ 

The suggestions of a classical-lag basis for the magnetic Aharonov-Bohm phase
shift were often rejected by the referees and editors of the leading physics
journals. \ Neither the referees nor the editors could see a classical
electromagnetic basis for a force on the passing electrons; and until such a
force was identified, all lag suggestions were unacceptable. \ There were
hints at the basis for a lag-producing force, but the solenoid was a
multi-particle system which seemed very hard to analyze.\cite{B2006} \ Lag
suggestions were published in journals which were more tolerant of views
outside the main stream. \ 

\subsection{Search for a Macroscopic Classical Lag}

In 2007, Caprez, Barwick, and Batelaan\cite{CBB} looked for a classical lag
effect using electrons passing macroscopic magnets, and they could find no
effect. \ However, it was suggested that the failure of this experimental
search might involve a large value of the solenoid resistance which negates
the energy balance ideas of the lag analysis.\cite{B2006a} \ Significant
electrical resistance in the windings of a macroscopic solenoid could lead to
energy absorption which would be larger than the changes in electromagnetic
energy used to obtain the lag result in Eq. (\ref{lag}). \ Although the lag
explanation offered here fits with the interference pattern shifts observed
for micro solenoids and toroids, no time delay has yet been measured using a
macroscopic solenoid or toroid. \ 

\subsection{Related Charge-Magnet Problems}

\ In 1967, Shockley and James\cite{SJ} had proposed a paradox involving
magnets and charges, which was addressed by Coleman and van Vleck,\cite{CV}%
\cite{B2015c} \ Coleman and van Vleck concluded that there was hidden momentum
in a magnet when there was a stationary charge outside. \ In a footnote, they
gave an example of mechanical hidden momentum. \ 

In 1984, Aharonov and Casher\cite{AC} suggested what they claimed was an
\textquotedblleft dual\textquotedblright\ of the magnetic phase shift. \ They
swapped positions of the charge and the magnetic moments. \ They proposed that
a phase shift would still be observed when the passing charge was extended to
a line charge and the solenoid was shrunk to a single magnetic dipole. \ The
phase shift for this Aharonov-Casher effect was experimentally
observed\cite{COW} in 1989. \ The model for the magnetic dipole used by
Aharonov and Casher involved two \textit{magnetic charges}, which indeed
experienced no net classical force when passing the line charge; however, when
the magnetic moment was modeled as a current loop, there was an obvious
classical force on the magnetic moment which would account for the observed
phase shift as a classical lag effect.\cite{B1987b} \ Aharonov, Pearl, and
Vaidman\cite{APV} agreed that there was a classical force on the magnetic
moment treated as a current loop, but contended that the magnetic moment still
moved with constant velocity (as through there were no force) because of
\textquotedblleft hidden mechanical momentum in magnets.\textquotedblright%
\ \ And so the idea of \textquotedblleft hidden mechanical momentum in
magnets\textquotedblright\ entered prominently into the physics
literature.\cite{hid}

In 2012, Mansuripur\cite{Mans} claimed to present \textquotedblleft
incontrovertible theoretical evidence of the incompatibility of the Lorentz
[force] law with the fundamental tenets of special
relativity.\textquotedblright\ \ Once again, the interaction of a point charge
and a magnet was involved, and questions involving conservation of angular
momentum were raised. \ There were at least a half-dozen replies to
Mansuripur's assertion, most invoking the idea of hidden mechanical
momentum.\cite{Bcontr}

\subsection{Separating Transient and Steady-State Problems}

It is clear that the interactions between charged particles and magnets
involves a complicated problem which has troubled physicists for many years.
\ It is the opinion of the present writer that the literature of the
Aharonov-Bohm phase shift and of the related paradoxes is full of
inaccuracies. \ In particular, the idea that \textquotedblleft hidden
\textit{mechanical} momentum in magnets\textquotedblright\ should be important
in an electromagnetic system with many closely-spaced charges seems absurd.
\ One should recall that the mechanical energy of the current-carriers of a
solenoid is miniscule compared to the large electromagnetic energy. \ Indeed,
in the texts of classical electromagnetism, the tiny mechanical kinetic energy
of the current carriers goes entirely unmentioned compared to the magnetic
energy in the solenoid. \ 

It seems important to separate out the transient interactions involved in the
Aharonov-Bohm and Aharonov-Casher phase shifts from the essentially
steady-state situations in the Shockley-James paradox, \textquotedblleft
hidden mechanical momentum\textquotedblright\ ideas, and the Mansuripur
suggestion. \ There can be internal momentum in magnets for steady-state
situations, but it is associated with electromagnetic fields, and not with
\textquotedblleft mechanical\textquotedblright\ momentum which is tiny
compared with electromagnetic momentum.\cite{B2015c} \ \ 

On the other hand, the transient effect in the Aharonov-Bohm situation seems
somewhat analogous to a transient in an $L/R$ circuit where the value of \ the
self-inductance $L$ is large. \ The essential aspect involves the back
(Faraday) acceleration fields which tend to oppose the effort to change the
speeds of the current carriers of the magnet. \ If the transient is of short
duration $\Delta t<<L/R$, then the current carriers accelerate briefly to give
the back (Faraday) electric fields which balance the magnetic energy changes,
but the currents associated with the magnet suffer only small changes. \ The
unfamiliar aspect in the Aharonov-Bohm situation is the \textit{localization}
at a single charge $e$ of the agent causing the acceleration of the current
carriers. In this situation without high symmetry, it is difficult to apply
the familiar calculations for changing magnetic fields in highly-symmetric
solenoids. \ If the ideas of the present article are indeed correct regarding
the basis for the phase shift, they suggest further experimental tests
involving varying\ both the solenoid and the $L/R$ ratio used in the
Aharonov-Bohm interference-pattern shift. \ 

\section{\ Conclusions}

In this article, we once again suggest the possibility that the
experimentally-observed Aharonov-Bohm phase shift arises because of a
classical electromagnetic interaction between the charge carriers of the
solenoid and the electromagnetic fields of the passing charges. \ Here we
emphasize the back (Faraday) acceleration electric fields which tend to
prevent the change in motion of groups of closely-spaced charged particles and
which place forces upon the agent trying to change the speed of the current
carriers. \ We first illustrate the large-electromagnetic-inertia effect for a
simple ring of point charges interacting with a single charge $e$. \ Next we
consider a ring of electric dipoles. \ Finally, we calculated the classical
lag of a charge passing through a magnetic toroid. \ The (many) charge
carriers of the magnetic toroid experience (very small) accelerations which
involve only small changes in speed while giving rise to \ a back (Faraday)
acceleration field which acts on the passing charge. \ The classical
electrodynamic analysis here predicts an Aharonov-Bohm phase shift where the
magnitude of the angle of the deflection of the double-slit pattern is the
same as that given by quantum theory. \ However, the direction of the
deflection, while given unambiguously by the classical analysis here, may
differ from that of quantum theory or of experiment.

\bigskip

ConstantCurrents7.tex \ \ \ \ \ \ \ \ \ \ \ \ \ \ \ \ \ December 10, 2022

\end{document}